# Improving HTc Josephson Junctions (HTc JJ) by annealing: the role of vacancy-interstitial annihilation.

M. Sirena, S. Matzen, N. Bergeal and J. Lesueur.

*Laboratoire Photons Et Matière, ESPCI, 10 Rue Vauquelin 75231 Paris (France).*

G. Faini

*Laboratoire de Photonique et Nanostructures, Route de Nozay, 91460 Marcoussis (France).*

R. Bernard, J. Briatico

*UMR-CNRS/THALES, Route D128, 91767 Palaiseau (France).*

Abstract.

We have studied the annealing effect in transport properties of High temperature Josephson Junctions (HTc JJ) made by ion irradiation. Low temperature annealing (80°C) increases the JJ transition temperature ($T_J$) and the Ic.Rn product, where Ic is the critical current and Rn the normal resistance. We found that the spread in JJ characteristics can be lowered by sufficient long annealing times. Using random walk numerical simulations, we showed that the characteristic annealing time and the evolution of the spread in JJ characteristics can be explained by a vacancy-interstitial annihilation process rather than by an oxygen diffusion one.

In the last fifteen years or so, the research on High Tc Josephson Junctions (HTc JJ) has been very active[1], as explained by their important potentialities for advanced electronics. The last ITRS roadmap[2] clearly indicates that JJ may be an alternative "beyond CMOS" for high speed electronics, and the recent SCENET roadmap on superconducting electronics[3] draws the detailed lines of the effort needed to be done for JJ to be a key technology in electronics and detection systems. Most of these applications require close packed arrays of junctions, whose characteristics (critical current Ic and current density Jc, normal state resistance Rn) have to be uniform over large areas and on large scales (from a few to tens of thousands of junctions)[4]. The recently developed fabrication technique of HTc JJ by ion irradiation appears very promising in this respect[5], and the authors have recently achieved a very good reproducibility of high quality JJ[6-8], thanks to the control and optimization of the fabrication process[6]. To go one step beyond and improve further the homogeneity of JJ characteristics, we propose in this paper a new tool : low temperature annealing. Annealing has been used to homogenize the depth defects profile in multi-energy irradiated JJ in $YBa_2Cu_3O_{7-\delta}$ (YBCO) films[9], and to increase their long term stability[10]. We present for the first time that it can lower the spread in characteristics within a JJ array.

We have measured the transport properties of several HTc JJ made in a 150 nm thick c-axis oriented YBCO film grown on $SrTiO_3$ single crystal by Pulsed Laser Deposition. Details of the JJ fabrication are given elsewhere[6,11]. The shadow mask used to define the junctions is made of polymethyl methacrylate photoresist in which 20 nm wide slits are opened by electron beam lithography. The samples were irradiated with 100 keV and 150 keV oxygen ions ($O^+$) with doses $\phi^{100keV} = 5.5 \times 10^{13}$ ions/cm$^2$ and $\phi^{150keV} = 6.75 \times 10^{13}$ ions/cm$^2$, chosen to induce the same defects density in the center of the barrier, i.e. the same corresponding critical temperature Tc'. The samples were annealed at 80°C with increasing annealing times t, ranging from 10' to 3600', and measured in a cryostat between 4K and 300K by standard four probes technique. Below the transition temperature of the reservoirs (90K), and above Tc', a Josephson behavior is observed up to a coupling temperature $T_J$[5]. In this temperature range, the I-V characteristics display a Resistively Shunted Junction (RSJ) like behavior[12], and the critical current follows a quadratic temperature dependence in agreement with the de-Gennes Wertamer model[13].

Figure 1 shows the normalized resistance as a function of temperature T for 100 keV irradiated JJ at different annealing times (10', 90', 600' and 3600'). Clearly $T_J$ increases upon annealing, while Rn decreases (not shown here). When plotting the average value $<T_J>$ as a function of time t, a logarithmic dependence is observed (inset Figure 2), which indicates that an activated process is involved. A very important result for technological applications, is that the width of the distribution of $T_J$ ($\Delta T_J$) first increases, and then *decreases* for long annealing time. We will return to this result later. Correspondingly, the critical current Ic increases (see Figure 2) and so does the IcRn product (the Josephson coupling energy) as seen in the inset Figure 2 on a logarithmic scale.

Those observations are compatible with a decrease of the defect density within the barrier: rather intuitively, low temperature annealing cures defects, and therefore restores better superconducting properties. Oxygen diffusion has been proposed to explain this behavior[9], in the context of multi-energy irradiation scheme. In that case the model considers diffusion along a direction perpendicular to the superconducting layer. To account for the experimental data, the authors found an activation energy for oxygen diffusion $E_D$ below the lowest one found in the literature (0.2 eV, as compare to 0.3 eV[14] and 1.2 eV[15]). We made an estimate of the characteristic oxygen diffusion time $t_D = \sigma^2/2D$ (where D is the diffusion coefficient and $\sigma$ the defects lateral distribution width ~ 70 nm[16]) in our annealing conditions. Given the wide spread of values for $E_D$ and D in the literature, we chose the values giving the highest characteristic time, corresponding to $D = D_0\exp(-E_D/k_BT)$ with $E_D$ = 0.33 eV and $D_0 = 6.25.10^{-9}$ cm$^2$/s[14]. For these values the calculated characteristic diffusion time is 3.4'. In that case we would not expect much change of the JJ characteristics after 30 minutes annealing, as opposed to the above presented data, where the evolution is still sizeable for 90 minutes and even three days. We therefore turned to another mechanism to explain our results: the vacancy-interstitial annihilation process. The defects produced by ion irradiation are mainly vacancy-interstitial Frenkel pairs. Upon annealing, they recombine according to a thermal activated process, curing the effects of defects. This has been extensively studied in semiconductors, Si for instance[17]. Since this process does not require long range diffusion, it can be efficient at rather low temperature, and may also explain the results of Tinchev et al[9].

To test the validity of this hypothesis, we made random walk type of simulations based on both mechanisms: diffusion and vacancy-interstitial annihilation. In the first case, the simulation mimics the diffusion of defects with constant hopping probability. In the second case, annihilation occurs with a probability depending on the square of the defect concentration, since both vacancy and interstitial have to be present for the Frenkel pair to disappear. More details about those simulations will be published shortly[18]. It is worth noticing that we use the *actual* defect distribution[16] given by the TRIM code[19], and not a pure Gaussian one, and that all process takes place along the barrier and not in the deepness of the film. The most stringent test for distinguishing between the two mechanisms is to account for the peculiar behavior of $\Delta T_J$. In the simulations, the local defect density lowers Tc' through an Abrikosov-Gorkov law[20] and therefore $T_J$. The spread in $T_J$ follows the spread in Tc', as it has been shown previously[16]. Therefore, we can compute $\Delta T_J$ as a function of $T_J$ to eliminate all the specific numerical simulation parameters (i.e. the hopping probability). Careful tests of the validity of this procedure have been made[18]. Figure 3 displays the results of the simulations (diffusion dotted line and annihilation solid line) and the experimental data for both 100 and 150 keV irradiated samples. The only adjustable parameter is the initial homogeneity of the JJ, which controls the shape of the curves. The values of the initial defects distribution dispersion used for the fitting are 3% for the JJ irradiated at 150 keV and 7% for the ones irradiated at 100 keV which are coherent with the experimental data ($\Delta T_j^0$ = 1.1K and 3.1K respectively). This result can be explained considering the slits size spread as the source of irradiated JJ $\Delta T_j$, as it was shown in a previous work[16]. The variation's amplitude of the variation has been adjusted for the simulations to account for the experimental one. In no case the diffusion process is able to reproduce the data, whereas the annihilation one does. The reason is rather fundamental. The initial increase of $\Delta T_J$ in both cases is due to the stochastic nature of the processes. The evolution of the number of defects in the center of the barrier ($N_i$) in each JJ is not univocally determined, but governed by a *distribution of probabilities of* $N_i$ ($P(N_i)$). This later is characterized by a mean value, which gives the mean Tc' (and therefore $T_J$) and a characteristic distribution width $\sigma_N$ which essentially gives $\Delta Tc'$ ($\Delta T_J$). Using the numerical simulations we have studied the evolution of $P(N_i)$ as a function of the annealing time, which depends on the physical mechanism. Diffusion is a very efficient process, and $\sigma_N$ starts decreasing for short annealing times. However, the vacancy-interstitial annihilation mechanism is less efficient since it does not present a "healing" or recovery mechanism once $N_i$ departs from the expected mean value. Further details about this will be given in 18.

Therefore we have evidence a mechanism which explains the behavior of the irradiated junctions upon annealing. Frenkel pairs annihilation restore junctions properties corresponding to lower irradiation dose $\Phi$. When $T_J$ goes up logarithmically, so does the IcRn product evaluated at Tc' (see inset Figure 2), as already observed by changing $\Phi$[11] ; correspondingly, the slope dIc/dT increases as expected. When $\Delta T_J$ increases, so does the spread in critical current (Figure 2). Moreover, this mechanism can improve the reproducibility of an array of junctions on a wafer and in some conditions, the spread in $T_J$ after annealing can be lower than the initial one (cf figure 3). When annealing at 80°C for an hour or so, the stability of the junctions is improved: further annealing at lower temperature (50°C) does not modify their characteristics.

In conclusion, we have shown that low temperature (80°C) annealing of irradiated HTc JJ restore their properties as if the effective defect density is lowered. Through experiments and random walk simulations, vacancy-interstitial annihilation has been proved to be the main mechanism governing the evolution of HTc JJ. Long (10 hours) annealing can decrease the spread of their characteristics within a wafer, and improve their thermal stability. These elements appear to be very encouraging for future applications.

The authors gratefully acknowledge O. Kaitasov for the ion irradiation made at IRMA-CSNSM (Orsay France), E Jacquet for the film growth and the personnel of the LPN-CNRS clean room for extraordinary technical support. M. Sirena acknowledges financial support from the MPPU- CNRS department.

[1] J. Halbriter, Superconductor Science & Technology **16** (10) R47 (2003). (1)

[2] International Technology Roadmap for Semiconductors: http://www.itrs.net/reports.html.

[3] H.J.M ter Brake et. al., Physica C, **439**, 1 (2006). (3)

[4] S. A. Cybart, K. Chen, Y. Cui, Qi Li, X. X. Xi, and R. C. Dynes, Appl. Phys. Lett **88**, 012509 (2006).

[5] S. S. Tinchev, Physica C, in press, doi: 10.106/j.physc.2007.04.001

[6] N. Bergeal, X. Grison, J. Lesueur, G. Faini, M. Aprili, and J. P. Contour, Appl. Phys. Lett. **87**, 102502 (2005).

[7] N. Bergeal, J. Lesueur, G. Faini, M. Aprili, J.P. Contour, Appl. Phys. Lett. **89**, 112515 (2006) and N. Bergeal, M. Sirena, J. Lesueur, G. Faini, M. Aprili, J.P. Contour, Appl. Phys. Lett. **90**, 136102 (2007).

[8] J. Lesueur, N. Bergeal, M. Sirena, X. Grison, G. Faini, M. Aprili, and J.P. Contour, IEEE Transactions on Applied Superconductivity (2007).

[9] U. Barkow, D. Menzel and S.S. Tinchev, Physica C, **370**, 246 (2002).

[10] F. Kahlmann, A. Engelhardt, J. Schubert, W. Zander, Ch. Buchal and J. Hollkott, IEEE Transactions on Applied Suprecond. **9**, 2874 (1999).

[11] N. Bergeal, J. Lesueur, M. Sirena, G. Faini, M. Aprili, J-P. Contour, B. Leridon, Journal of Appl. Physics (accepted for publication).

[12] A. Barone and G. Paterno, Physics and applications of the Josephson effect (Wiley, New York, 1982).

[13] P. G. Degennes and E. Guyon, Physics Letters **3,** 168-169 (1963).

[14] S.H. Lee, S.C. Bae, J.K. Ku, H.J Shin, Phys. Rev. B., **46**, 9142 (1992).

[15] Y. Li, J.A. Kilner, T.J. Tate, M.J. Lee, R.J. Charter, H. Fox. R.A. De Souza, P.G. Quincey, Pjys. Rev. B. **51**, 8498 (1995).

[16] M. Sirena, N. Bergeal, J. Lesueur, R. Bernard, J. Briatico, D. Crété, Journal of Appl. Physics, **101**, 123925 (2007).

[17] S. Libertino, et. al. Appl. Phys. Lett. **71**, 389 (1997).

[18] M. Sirena, et. al., to be published.

[19] J. F. Ziegler, and J. P. Biersack, SRIM (IBM, New York, 2004).

[20] J. Lesueur, L. Dumoulin, S. Quillet and J. Radcliffe, Journal of alloys and compounds **195**, 527 (1993).

Figure 1: Normalized resistance (Rn) as a function of temperature (T) for JJ made on the same wafer by 100 keV oxygen ions irradiation, and then further annealed for different time (t=10, t=90', t=600' and 3600'). The arrows indicate the maximum spread in $T_J$ for a given annealing time.

Figure 2: Critical current as a function of the reduced temperature for the Josephson Junctions irradiated at 100 keV for different annealing times (t=10', t=90, and t=600'). Lines between the curves are a guide for the eyes. Arrows indicate the mean value and the spread of Ic at Tc'. The insert shows Tj and the Ic.Rn product for the JJ irradiated at 100 keV as function of the annealing time.

Figure 3: $T_J$' spread ($\Delta T_J$) as a function of $T_J$ for the JJ made on the same wafer by 100 keV (closed symbols) and 150keV (open symbols) oxygen ions irradiation. Shown in solid line are the simulation results for the vacancy-interstitial annihilation model, and in dashed lines for the diffusion mechanism.

Figure 1 : Sirena et. al.

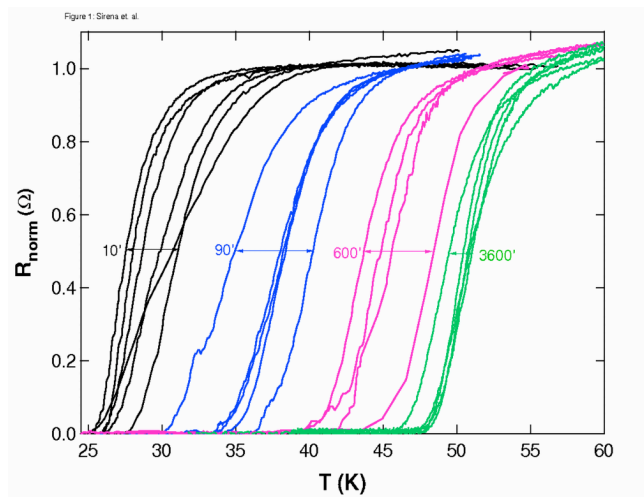

Figura 2: Sirena et. al

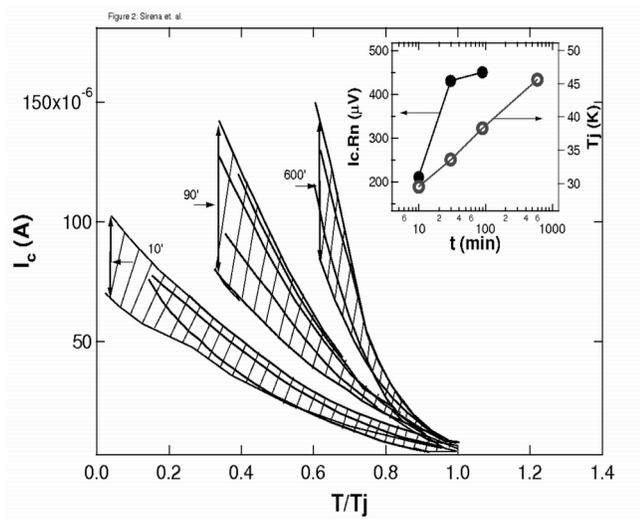

Figura 3: Sirena et. al.

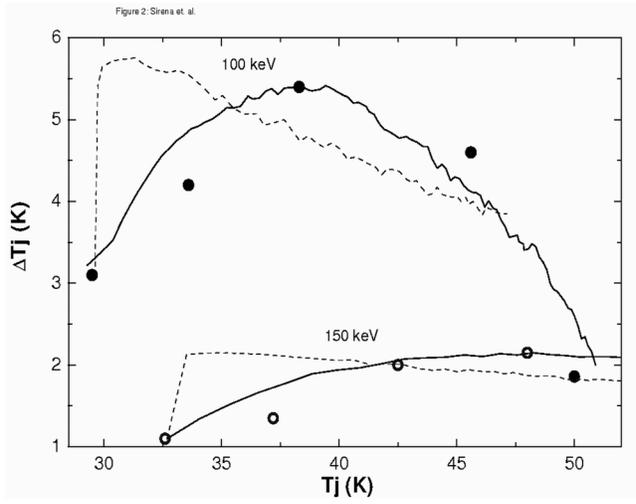